\documentclass[runningheads]{llncs}
\usepackage{graphicx} 
\usepackage{caption}
\usepackage{subcaption}
\usepackage[dvipsnames]{xcolor}

\definecolor{blue}{rgb}{0, 0, 1}

\begin{document}
\title{Neural Spherical Harmonics for structurally coherent continuous representation of diffusion MRI signal}
\titlerunning{Neural spherical harmonics}
%
\author{Tom Hendriks\inst{1}\orcidID{0000-0001-6374-4531}\and
Anna Vilanova\inst{1}\orcidID{0000-0002-1034-737X} \and
Maxime Chamberland\inst{1}\orcidID{0000-0001-7064-0984}}
\authorrunning{T. Hendriks et al.}
%
\institute{Eindhoven University of Technology, Department of Computer Science and Mathmatics, Groene Loper 5, 5612 AP Eindhoven, The Netherlands\\
Corresponding author: \email{t.hendriks@tue.nl}}
\maketitle              

\begin{abstract}
We present a novel way to model diffusion magnetic resonance imaging (dMRI) datasets, that benefits from the structural coherence of the human brain while only using data from a single subject. Current methods model the dMRI signal in individual voxels, disregarding the intervoxel coherence that is present. We use a neural network to parameterize a spherical harmonics\index{spherical harmonics} series (NeSH) to represent the dMRI signal of a single subject from the Human Connectome Project dataset, continuous in both the angular and spatial domain. The reconstructed dMRI signal using this method shows a more structurally coherent representation of the data. Noise in gradient images is removed and the fiber orientation distribution\index{fiber orientation distributions} functions show a smooth change in direction along a fiber tract. We showcase how the reconstruction can be used to calculate mean diffusivity\index{mean diffusivity}, fractional anisotropy\index{fractional anisotropy}, and total apparent fiber density\index{apparent fiber density}. These results can be achieved with a single model architecture, tuning only one hyperparameter. In this paper we also demonstrate how upsampling\index{upsampling} in both the angular and spatial domain yields reconstructions that are on par or better than existing methods.

\end{abstract}

\section{Introduction}
The human brain is a highly structured organ. With the introduction of diffusion magnetic resonance imaging (dMRI) in vivo study of the structure of the brain became a possibility. The spatially coherent structures in the brain imply that spatial coherence should be present when modeling dMRI data. Diffusion tensor imaging (DTI) \cite{basser1994estimation} fits a tensor for every voxel of the volume describing the diffusion in three primary directions. Constrained spherical deconvolution (CSD) \cite{tournier2007robust} can describe the orientation and relative size of fiber bundles using fiber orientation distribution functions (fODFs). These are examples of methods that model the fiber orientation in every voxel independently, disregarding any intervoxel coherence. Interpolating correctly between voxels using classical interpolation methods (e.g. cubic interpolation) is, therefore, difficult and susceptible to noise, and can discard anatomical details. Interpolation in the angular domain has proven to be a difficult task as well, as highlighted by recent challenges in the computational dMRI community \cite{bonet2019computationalgranada,tax2019cross,ning2020cross}. Machine learning\index{machine learning} approaches for upsampling in both the angular and spatial domains are a promising avenue \cite{ajavalidation}. However, these methods often rely on a strong prior obtained by training on large amount of data. This is problematic when training data is scarce or if the model is applied to data inherently different from the data it was trained on (e.g. pathological data). Ideally, a continuous\index{continuous representation} and structurally coherent\index{structural coherence} model should be derived at the individual level (i.e., n=1).\\
Neural radiance field\index{neural radiance field} (NeRF) \cite{mildenhall2021nerf} models have shown to be extremely effective at creating continuous 3-dimensional representations, known as implicit neural representations\index{implicit neural representation}, of scenes given a limited number of 2-dimensional input images taken from limited angles. NeRF overfits a multi-layer perceptron to essentially capture a given scene in its parameters. Unseen angles can then be sampled from this network. This concept could translate well to dMRI, as we are trying to create a complete representation from an incompletely sampled angular domain. The difference with dMRI is that every angle in a dMRI-acquisition produces a complete 3-dimensional volume of data.\\ 
In this work, we propose to use a NeRF-like model to create a model of the dMRI data of a single subject that utilizes the structural coherence of the brain, while providing continuity in both the angular and spatial domain. We evaluate the resulting model in a number of downstream tasks, such as calculating microstructural metrics\index{microstructural metrics}, and fODF estimation. We also demonstrate how the model can be used to upsample\index{upsampling} dMRI data in both the angular and spatial domain.

\section{Methods \& Experiments}
\subsection{Data}
We sourced data from a single participant from the preprocessed Human Connectome Project dataset \cite{van2013wu} consisting of 18 $b = 0 s/mm^2$ volumes, 90 $b=1000$ $s/mm^2$ volumes, 90 $b=3000$ $s/mm^2$ volumes, with 1.25mm isotropic voxels. 

\subsection{Model}
The neural spherical harmonics model (NeSH) is an adaptation from SH-NeRF \cite{yu2021plenoctrees}. NeSH outputs an approximation $\hat{S}(x, y, z, \vec{b})$ for a diffusion signal $S(x, y, z, \vec{b})$.
An input pair $i \in I$ consists of a voxel midpoint coordinate $(x, y, z)$ and a gradient direction vectors $\vec{b}$, where $I$ is a set of all possible coordinate-direction pairs. I has size $N = n_c \times n_d$ with $n_c$ being the number of coordinates and $n_d$ the number of directions. The input coordinates are scaled to lie in $[-1, 1]$ and are positionally encoded using the generalization of the NeRF positional encoding \cite{tancik2020fourier} into input vector $\vec{x}$.
Direction vector $\vec{b}$ is converted into the corresponding polar angles $\theta$ (azimuth, $[0, 2\pi)$) and $\phi$ (elevation, $[0, \pi]$).

\begin{figure}
    \centering\includegraphics[width=0.85\textwidth]{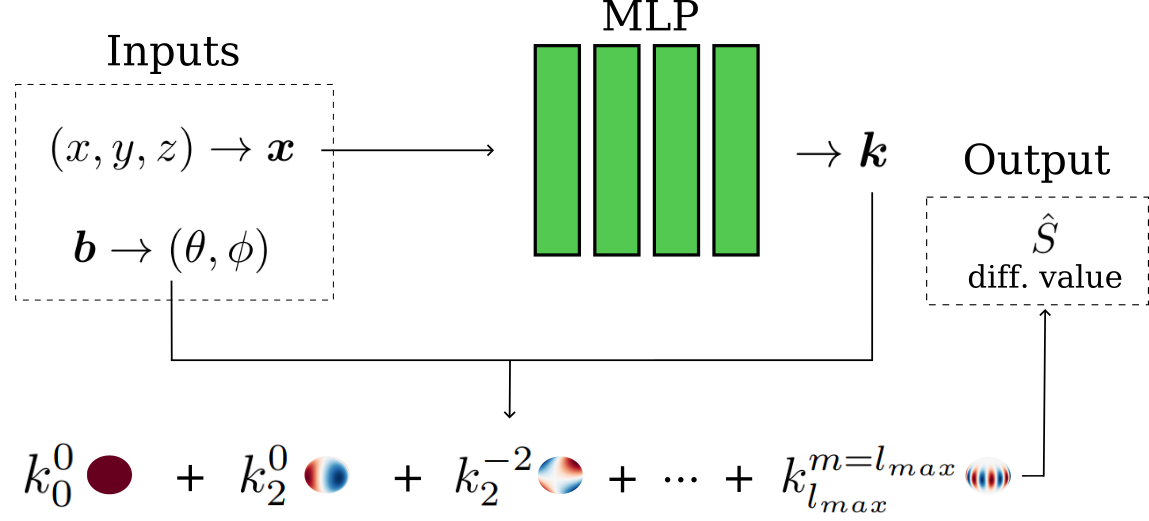}
    \caption{A schematic representation of the Neural spherical harmonics (NeSH) model. Inputs coordinates are spatially encoded into $\vec(x)$, directional vector $\vec{b}$ is converted to $\theta$ and $\phi$. Vector $\vec{x}$ is passed through the multi-layer perceptron (MLP) to produce $\vec{k}$, which parameterizes the spherical harmonics series. This is sampled in direction $\vec{b}$ to produce the final output $\hat{S}$.}
    \label{model_img}
\end{figure}

A simple multi-layer perceptron\index{multi-layer perceptron} (MLP) maps $\vec{x}$ into a coefficient vector $\vec{k}$ that parameterizes a spherical harmonics\index{spherical harmonics} (SH) series.
\begin{equation}
    M_\Psi: \vec{x} \rightarrow \vec{k}
    \label{MLP}
\end{equation} 
\begin{equation}
    \vec{k} = (k^m_l)^{m:-l\leq m \leq l}_{l:0\leq l \leq l_{max}}
\end{equation}
where $k^m_l$ is the coefficient for the SH component of degree $l$ and order $m$, $l_{max}$ is the maximum degree of the SH-series, and m the order. For a given $i$ we can now obtain an estimation of the dMRI signal: 
\begin{equation}
\label{sh_series}
    \hat{S_i} = \hat{S}(x_i, y_i, z_i, \vec{b_i}) = \sum_{(k^m_l)^i \in \vec{k_i}} (k^m_l)^iY^m_l(\theta, \phi) 
\end{equation}
where $Y^m_l(\theta, \phi)$ is the SH component of degree $l$ and order $m$ for azimuth $\theta$ and elevation $\phi$ obtained from $\vec{b_i}$, $\vec{k_i}$ is the coefficient vector given by (\ref{MLP}) for input $i$, and $(k_l^m)^i$ is the coefficient of $Y^m_l$. The dMRI signal is reconstructed with a simplified real basis SH series, using only odd-numbered degrees. Different methods of simplifying the SH series exist \cite{tournier2007robust,descoteaux2007regularized}; in this paper, the method described in MRtrix3 is used \cite{tournier2019mrtrix3}. The full model is shown in Figure \ref{model_img}.

We calculate the loss as an average over all inputs for the smooth L1 loss \cite{girshick2015fast} between the value of $\hat{S_i}$, and the dMRI signal $S_i$ defined as the dMRI signal measured at $(x_i, y_i, z_i)$ in the direction of $\vec{b_i}$. Unregularized, NeSH could be susceptible to overfitting on noise, if the maximum degree of the SH series is larger than necessary to model the diffusion data in a given voxel. An L1 regularization term is added as an incentive to minimize unnecessary coefficients. The resulting loss function is:
\begin{equation}
\label{loss_f}
    L = \frac{1}{N} \sum_{i\in I} \Bigl( smooth_{L1}(S_i - \hat{S}_i) + \lambda \bigl(\sum_{k^m_l \in \vec{k_i}} |(k_l^m)^i|\bigl) \Bigl)
\end{equation}
 where $|(k_l^m)^i|$ is the absolute value of the coefficient. The loss is used to update the MLP parameters $\Psi$.

To reconstruct images from the trained model, a set $C$ of $(x, y, z)$ coordinates is generated at the desired spatial resolution, as well as a set $B$ of directions in the desired angular resolution. A dMRI dataset is reconstructed by first positionally encoding, and mapping every coordinate $\vec{c}\in C$ to $\vec{k_c}$ using (\ref{MLP}), and then sampling the SH-series parameterized with $\vec{k_c}$ for every direction $\vec{b} \in B$. Effectively this applies (\ref{sh_series}) to every coordinate-direction pair, but only calculates $\vec{k_c}$ once for every input coordinate.\\
The model has the following hyperparameters: $l_{max}$ sets the maximum degree of the SH, $l_{pos}$ sets the number of positional encodings, $\sigma$ scales the positional encoding, $n\_layers$ sets the number of layers in the MLP, $hidden\_dim$ sets the number of neurons in each layer, $lr$ is the learning rate, $\lambda$ scales the L1 regularization.

\subsection{Implementation}
The model is implemented in python version 3.9.16, with pytorch version 2.0.0. MRtrix3 version 3.0.4 is used to calculate DTI metrics and fODFs, and to visualize results. Scilpy\footnote{https://github.com/scilus/scilpy} version 1.5 is used (with python version 3.10.10) to calculate fODF based metrics, and to create interpolated spherical functions. The BET of FSL version 6.0.6.4, is used for brain mask segmentation.

\subsection{Experiments}
\subsubsection{Reconstruction and angular upsampling of the dMRI signal}
To assess if the proposed model can reproduce the original data, NeSH is fit on 30 gradient directions of the $b=1000$ $s/mm^2$ shell. A grid-search is performed over the hyperparameters. Visual inspection of the gradient images, as well as DTI metrics\index{diffusion tensor imaging metrics} and fODF\index{fiber orientation distributions} glyphs, determine which settings produce the best results. Then, to assess if these settings can be applied to a different set of gradient directions, the settings found in part one are used to fit the model on 90, 60, 45, 30, 15, 10 and 3 gradient directions for both the $b = 1000$ $s/mm^2$ and $b = 3000$ $s/mm^2$ dMRI acquisitions. As a comparison, spherical harmonics interpolation\index{spherical harmonics interpolation} (SHI) \cite{descoteaux2007regularized} is fit on the same number of gradients. The root mean squared error (RMSE) is calculated for each of the models between the input gradient images and the reconstructed gradient images it produces:
\begin{equation}
\label{rmse_f}
    \sqrt{\frac{1}{WHD|B|}\sum^W_{x=1}\sum^H_{y=1}\sum^D_{z=1}\sum_{\vec{b}\in B} (S(x,y,z,\vec{b}) - \hat{S}(x,y,z,\vec{b}))^2}
\end{equation}
where $W$, $H$, and $D$ are the width, height, and depth of the image, $B$ is the set of gradient directions with size $|B|$, $S(x,y,z,\vec{b})$ is the measured signal, and $\hat{S}(x, y, z, \vec{b})$ is the reconstructed signal at location $x$, $y$, $z$ for gradient direction $\vec{b}$. Finally the capabilities of the model to upsample\index{angular upsampling} in the angular domain are assessed. The resulting models from the second part are sampled in all 90 gradient directions. The RMSE is calculated between the 90 original gradient images and the 90 reconstructed gradient images using (\ref{rmse_f}). In all experiments the RMSE is only calculated within a brain mask.

\subsubsection{Spatial upsampling}
The data modeled with NeSH can be sampled in any spatial resolution. This experiment assesses the quality of the data when upsampled\index{spatial upsampling} in spatial domain. The HCP dataset is downsampled from 1.25mm to 2.5mm isotropic voxels. NeSH is fit on the downsampled dataset using 90 gradient directions, and then sampled at 1.25mm isotropic resolution. The downsampled dataset is also upsampled to the original 1.25mm isotropic resolution using cubic interpolation. For the resulting datasets a color encoded FA map is calculated and visualized to compare the results.

\subsubsection{DTI and fODF Metrics}
In this experiment we assess if the data modeled with NeSH can be used to produce three common dMRI microstrutural metrics. Two DTI metrics: mean diffusivity\index{mean diffusivity} (MD) and fractional anisotropy\index{fractional anisotropy} (FA), and one fODF metric: total apparent fiber density\index{apparent fiber density} (AFD, \cite{raffelt2012apparent}). The metrics are calculated for 90 gradients, 90 gradients reconstructed by NeSH fit on 90 gradients, 30 gradients, 90 gradients reconstructed by NeSH fit on 30 gradients, and 90 gradients reconstructed by SHI fit on 30 gradients. The $b = 1000$ $s/mm^2$ shell was used for the DTI metrics, and the $b = 3000$ $s/mm^2$ shell for AFD. The three measures are compared to the ones obtained from the full 90 gradients set by computing and visualizing a difference map.

\subsubsection{fODF estimation}
This experiment is used to assess if fODFs\index{fiber orientation distributions} can be generated from data modeled with NeSH. The same datasets as in the previous experiment are used. A response function is first extracted from the dMRI acquisitions using the single shell implementation of the algorithm by Tournier \cite{tournier2013determination}. Secondly, the fODFs are calculated using single shell CSD \cite{tournier2007robust}. For all five datasets, the $b = 3000$ $s/mm^2$ shell is used. Results are visualized by showing fODF glyphs.

\section{Results}
\subsubsection{Reconstruction and angular upsampling of dMRI signal}
\begin{figure}
    \centering\includegraphics[width=0.85\textwidth]{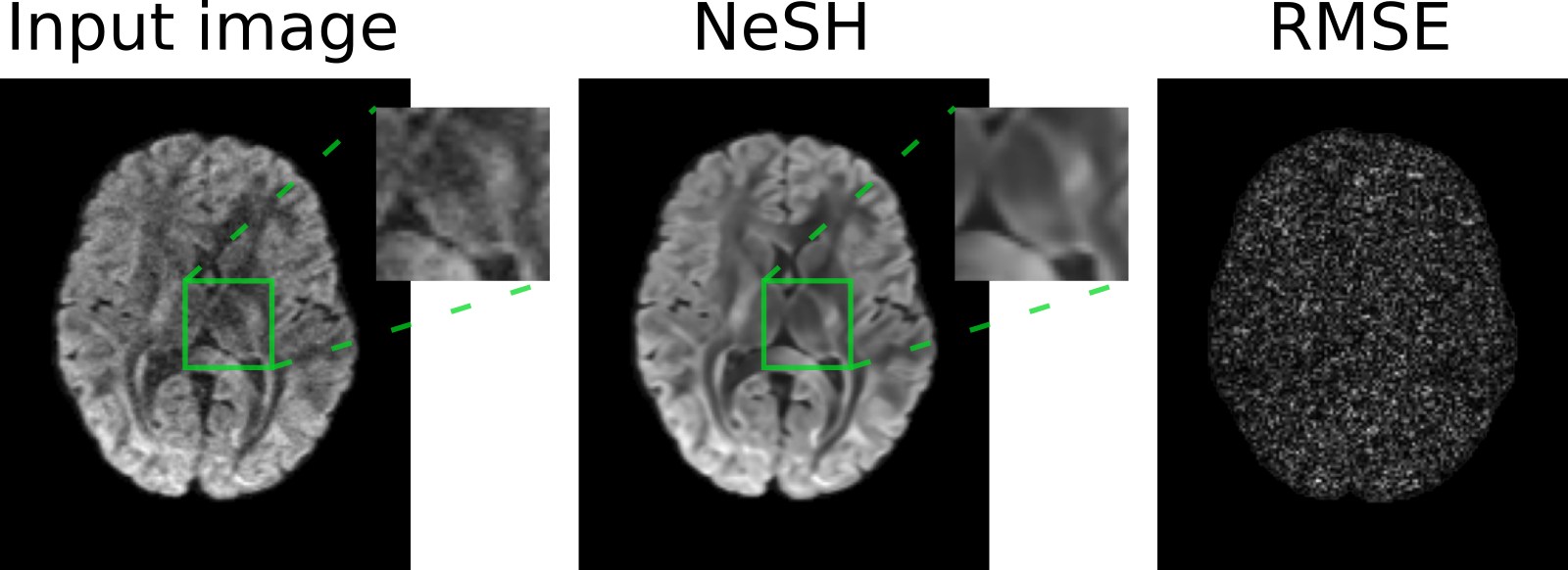}
    \caption{A single axial slice of dMRI data, from a single gradient direction shown as baseline (column one) and as reconstructed by NeSH (column two). The root mean squared error (RMSE) between the two images is shown in column three. A two time magnification of the area in the green boxes shows the denoising effect more clearly.}
    \label{grad_comparison_image}
\end{figure}

The grid-search over the parameters resulted in the following hyperparameter settings: $l_{max} = 8$ (for models trained $\leq10$ gradient directions $l_{max} = 2$), $l_{pos} = 12$ resulting in an input size of 75 (12 sine and cosine encodings for each dimension + raw coordinates), $\sigma = 4$, $n\_layers = 4$, $hidden\_dim = 2048$, $lr = 10\times10^{-5}$, $\lambda = 10\times10^{-6}$. The Adam optimizer was used with default settings. The model is trained for $5$ epochs with a batch size of $1000$. Figure \ref{grad_comparison_image} shows a slice of dMRI data for a single gradient direcion, the output generated by NeSH, as well as the mean squared error (RMSE) between the two images. NeSH produces a smoother image, removing the noise\index{denoising} from the input image. The noise appears to be randomly distributed, without anatomical residuals.

\begin{figure}
     \centering
    \begin{subfigure}[b]{0.49\textwidth}
         \centering
         \includegraphics[width=\textwidth]{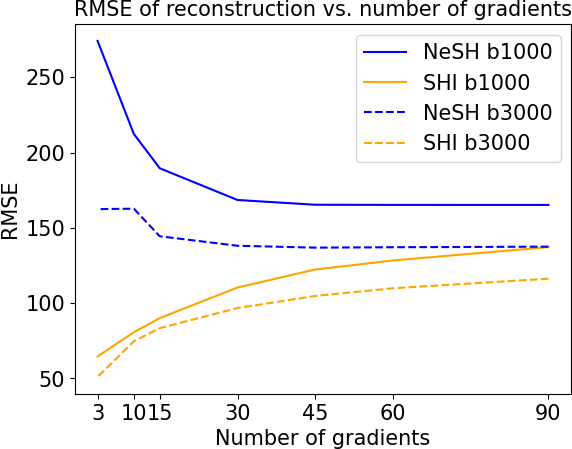}
         \caption{}
         \label{rmse_recon_img}
     \end{subfigure}
     \hfill
     \begin{subfigure}[b]{0.49\textwidth}
         \centering
         \includegraphics[width=\textwidth]{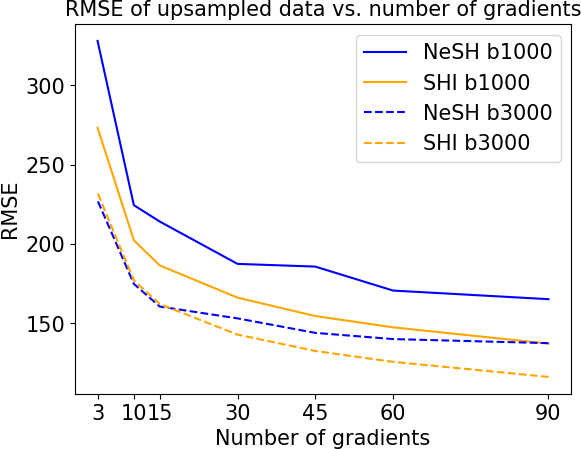}
         \caption{}
         \label{rmse_inter_img}
     \end{subfigure}
     \hfill
     \caption{Root mean squared error (RMSE) of the reconstructed dMRI volumes by NeSH and spherical harmonics interpolation (SHI) compared to the gradient images used to fit the model (a) and compared to the full set of 90 gradient images (b), for both $b = 1000$ $s/mm^2$ and $b = 3000$ $s/mm^2$.}
    \label{rmse_img}
\end{figure}

The comparisons of reconstruction error between NeSH and SHI are shown in Figure \ref{rmse_recon_img}. NeSH has a higher RMSE when reconstructing a lower number of gradients, which lowers with an increasing number of gradients. SHI has a lower RMSE when reconstructing a lower number of gradients, which increases with an increasing number of gradients. SHI has a consistently lower RMSE compared to NeSH on both $b = 1000$ $s/mm^2$ and $b = 3000$ $s/mm^2$.

The comparisons of upsampling error between NeSH and SHI are shown in Figure \ref{rmse_inter_img}. For both Nesh and SHI the RMSE of the upsampled data lowers when the model is fit on more gradient directions. SHI has a lower RMSE for all gradient subsets for $b = 1000$ $s/mm^2$, and for all subsets with more then 15 gradients for $b = 3000$ $s/mm^2$.

\subsubsection{Spatial upsampling}
\begin{figure}
\centering\includegraphics[width=0.6\textwidth]{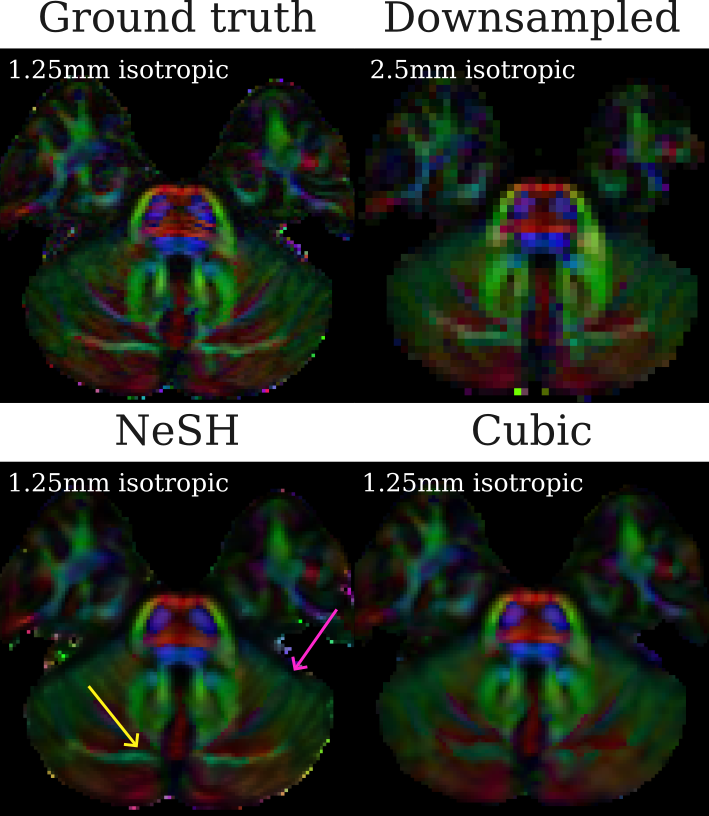}
\caption{A slice of color encoded FA maps at the cerebellar level, generated for different datasets. Clockwise starting top left: the 1.25mm isotropic original image (ground truth), the downsampled 2.5mm isotropic image, 1.25mm isotropic image upsampled using cubic interpolation, 1.25mm isotropic image upsampled using NeSH. The arrows show two areas where NeSH is able to reconstruct details of the cerebellum which are lost in cubic interpolation. } \label{rgb_img}
\end{figure}

Figure \ref{rgb_img} shows the color encoded FA\index{fractional anisotropy} maps for this experiment. The dataset reconstructed by NeSH fit on 90 gradients is able to reconstruct details in the cerebellar cortex and cerebellar white matter that are lost in cubic interpolation. Finer-grained details of the 1.25mm isotropic voxel data are lost in both upsampling methods.

\subsubsection{DTI Metrics and fODF metrics}
\begin{figure}
\includegraphics[width=\textwidth]{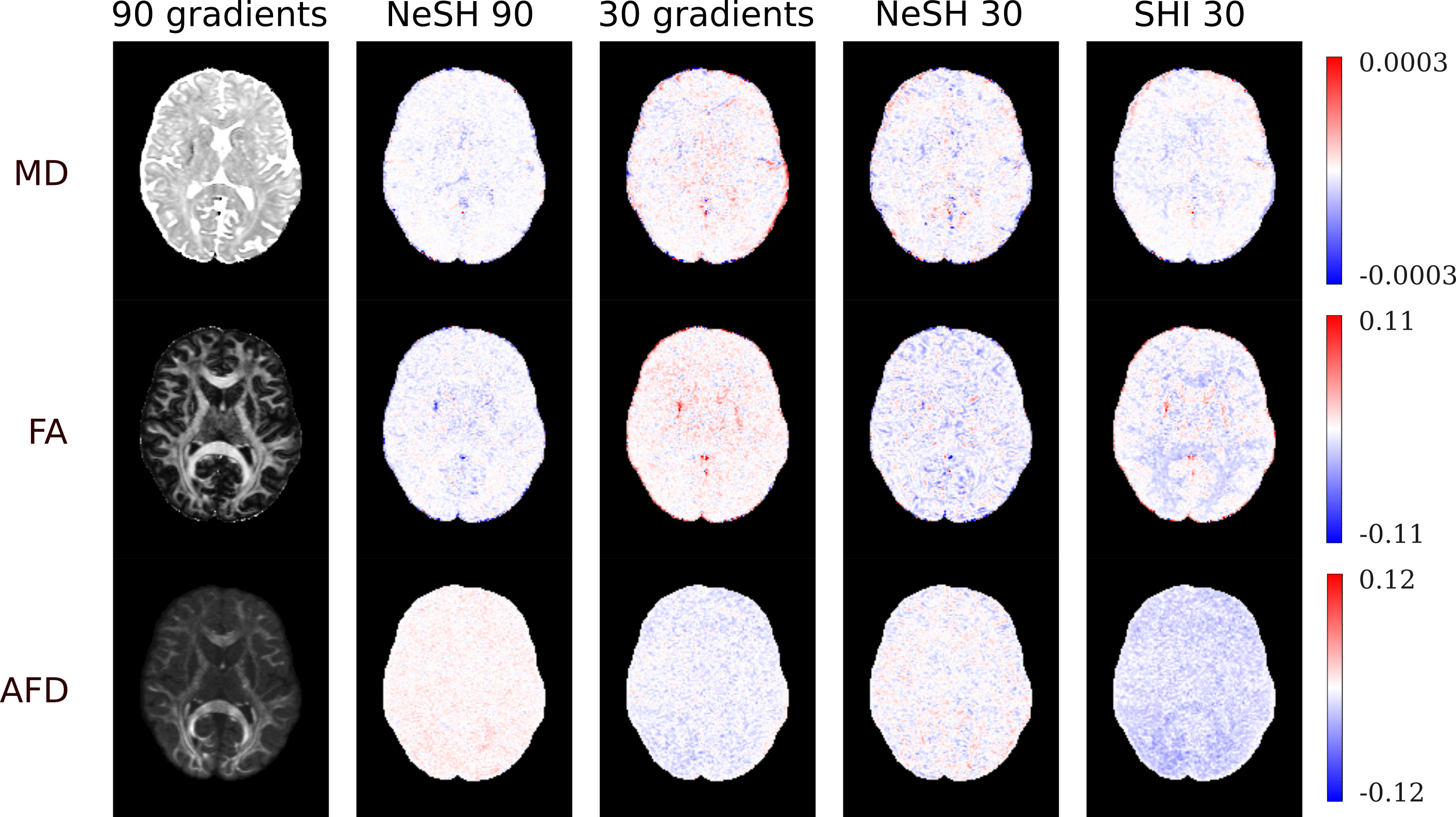}
\caption{Visualization of mean diffusivity (MD), fractional anisotropy (FA) and total apparant fiber density (AFD). In the first column the metrics are shown as a map for a single axial slice of the volume when calculated on the full set of 90 gradient images. The remaining columns show the difference map for the other datasets when compared to the 90 gradient images. Blue signifies negative difference, red signifies positive difference.} \label{dti_img}
\end{figure}

Figure \ref{dti_img} shows the results of this experiment. The DTI metrics (MD, FA)\index{mean diffusivity}\index{fractional anisotropy} show low error when downsampling, and in NeSH and SHI reconstructions on both 90 and 30 gradients, when compared to the metrics calculated on 90 gradients. Downsampling and SHI both over- and underestimate the metric, with most errors located in the white matter areas. NeSH more frequently underestimates the metrics and the errors are located more in the grey matter areas. The AFD\index{apparent fiber density} is overestimated by NeSH reconstructions, and underestimated when downsampling or resconstructing using SHI. In AFD the distribution of error is similar for all methods, but downsampling and SHI show an underestimation of the AFD, while NeSH shows errors in both directions.

\subsubsection{fODF estimation}
\begin{figure}
\includegraphics[width=\textwidth]{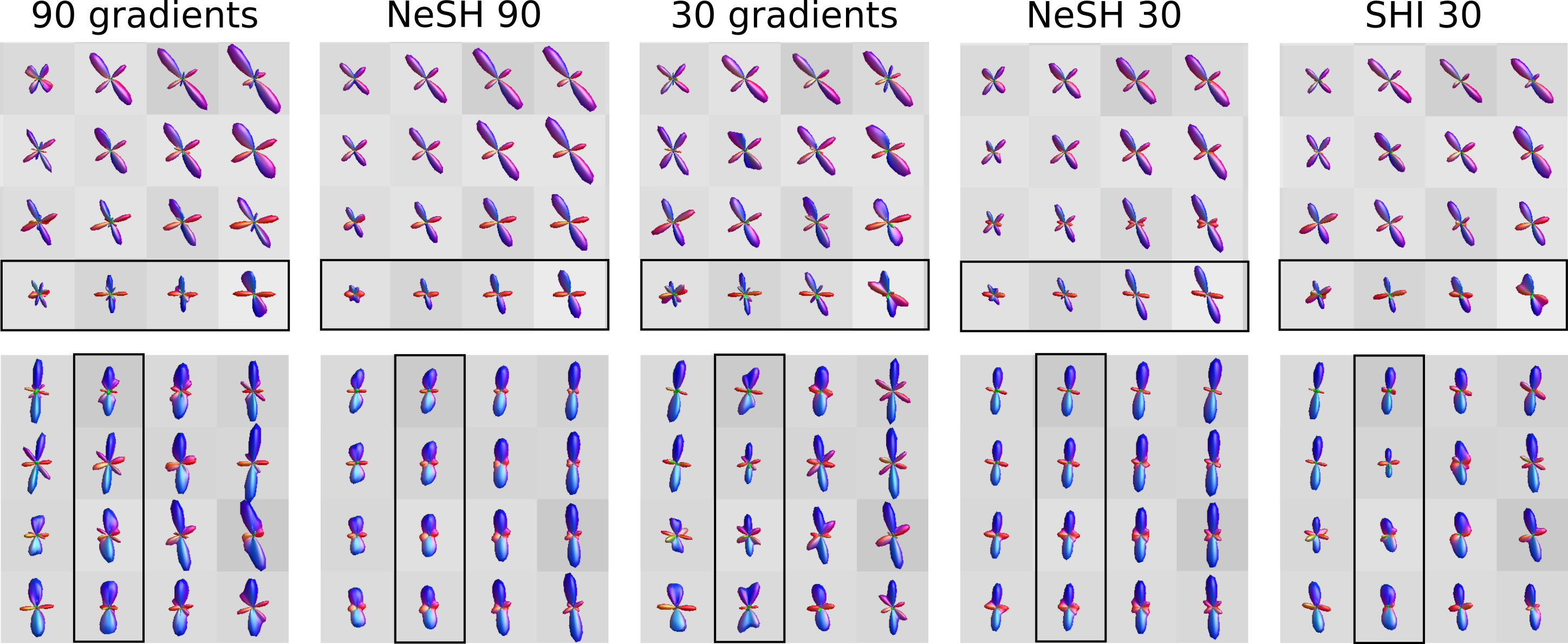}
\caption{Magnified coronal view showing the fiber orientation distributions glyps for the different datasets on a background of a T1-weighted image. First row: centrum semi-ovale, highlighted region shows the increased intervoxel consistency in NeSH modelled data. Second row: descending part of the corticospinal tract, highlighted region shows increased intervoxel consistency in NeSH modelled data, as well as a decrease in the size of the crossing fibers.} \label{fods_img}
\end{figure}

The visualization of a group of fODFs\index{fiber orientation distributions} in the centrum semi-ovale and the descending part of the CST can be seen in Figure \ref{fods_img}. The glyphs created from the NeSH reconstructions show a smooth, structurally-coherent change, while maintaining the important information, i.e. the crossing of fibers in the centrum semi-ovale. In presence of a big fiber tract such as the CST, the NeSH reconstruction shows a decrease in amplitude in other directions. In other methods, the fODFs exhibit more noise and less alignment between voxels, while the peaks appear sharper.

\section{Discussion}
We have introduced a novel method to model only a single acquisition of dMRI data using a neural representation\index{implicit neural representation} of spherical harmonics\index{spherical harmonics}, called NeSH. We show that dMRI data reconstructed by NeSH appears to be denoised\index{denoising} compared to the original data (Figure \ref{grad_comparison_image}. We hypothesise that the model is able to capture the continuous structures of the brain, but not the erratic nature of the noise. The RMSE lies consistently higher for both the reconstruction and upsampling compared to SHI (Figure \ref{rmse_img}, which could partly be explained by the removal of noise. For the reconstruction of the input gradients, NeSH performs worse with a decreasing number of input gradients. Possibly this indicates that to find a good representation NeSH needs a minimum number of gradients, which appears to be around 15. Both NeSH and SHI show an increase in RMSE when upsampling\index{angular upsampling} to 90 gradients from a decreasing number of gradients, as is to be expected.

We also show that NeSH can also be used to upsample\index{spatial upsampling} in the spatial domain. Figure \ref{rgb_img} shows that NeSH is able to reconstruct details that are not clearly visible in the 2.5mm isotropic data, but are present in the 1.25mm ground truth. This strengthens the hypothesis that using multiple gradient directions, NeSH can model a continuous representation of the dMRI data. While the achievable level of detail is lower than achieved by Alexander et al \cite{alexander2017image}, it does not rely on a prior learned from a large dataset.

Furthermore, we show that the reconstructed dMRI volumes can be used to calculate MD\index{mean diffusivity}, FA\index{fractional anisotropy}, and AFD\index{apparent fiber density}. Compared to the metrics calculated using 90 gradient direction, NeSH differs mostly in the gray matter areas, while downsampled volumes and SHI reconstructed volumes have differences in the white matter areas. This supports our hypothesis that NeSH benefits from the structural continuity of the fiber bundles to model the data. As the white matter areas are usually the areas of interest, this could be seens as a benefit of NeSH. The increased brightness in the posterior commissure and surrounding tissue can be explained by the lack of bias field correction in pre-processing.

Finally, we show that fODFs\index{fiber orientation distributions} generated from NeSH reconstructions have a smooth change in fiber directions between voxels. This is also supportive of the structural continuity hypothesis. The 90 gradient, 30 gradient, and SHI reconstructed FODs show a more erratic pattern, in which the FODs are less aligned overall. The decrease in size of the crossing fibers in the descending part of the CST shows that NeSH prioritizes the major bundle in this area. Some information on possible smaller bundles is now lost, however, which is something that should be looked into in future versions.

\subsubsection{Limitations} The lack of a gold-standard in dMRI complicates the interpretation of the experiments. The denoising effect shown in Figure \ref{grad_comparison_image} is an example. We cannot be certain if the representation modelled by NeSH is a more realistic one than the more noisy representation or just a smoother one. In the last experiment NeSH consistently shows lower values in grey matter areas. The fODFs in the grey matter area correspondingly show a less peaked, smaller amplitude. Compared to existing techniques this can be interpreted as an error, but anatomically it makes sense as there are no large fiber bundles in the grey matter. Additionally, with no ground truth data, it is difficult to assess how good the representation outside of the voxels actually is. Synthetic datasets with known ground truths can provide a better idea. Furthermore, both the architecture and positional encodings used in this paper are simple. Many developments in the field NeRF have taken place since \cite{zhu2023deep,tancik2020fourier}.  Architectural and methodological changes to NeSH could lead to further improvement. Finally, we choose to model the dMRI signal directly through an SH-series, in order to evaluate the data quality with a variety of downstream tasks. This is not a necessity. Anything that can be transformed into dMRI signal can be modelled by NeSH (e.g. peak directions or fODFs, which can be convolved into a diffusion weighted signal).

\subsubsection{Future work} 
Future work will further investigate the advantages of modelling dMRI data in a continuous space, as well as further evaluate the findings of the experiments. First, the quality and usability of the denoising properties of NeSH should be compared to other existing denoising methods. Second, using clinical datasets of lower angular and spatial resolution can provide insight into the 'real-world' clinical applicability of NeSH. This is especially interesting in MRI acquisitions of pathology (e.g. a glial-cell tumor) in the brain, as models relying on a prior learned on outside data might fail here. The harmonization of dMRI datasets across scanners and protocols \cite{tax2019cross,ning2020cross}, is another area of research where NeSH can be applied. Third, fiber tracking is a common use-case for dMRI. We have performed fibertracking using iFOD2 \cite{tournier2010improved} with tract masks and begin- and endpoint inclusion for different numbers of gradient directions. This showed no major differences between the different methods for all inspected tracts. An explanation for this is the high spatial resolution of the HCP data, which allows tracts to be generated even with a downsampled angular resolution. Further research on datasets with lower spatial resolution will have to show the value of using NeSH reconstructions for fiber tracking. Fourth, a recent paper by Mancini et al \cite{mancini2022lossy} has shown how compression of dMRI data using sinusoidal representation networks (SIRENs) \cite{sitzmann2020implicit} does not lead to reduced quality in downstream tasks. Using a SIREN architecture could also prove useful for the SH-based approach we have described. Finally, the generalization of the model to other subjects, protocols, and scanners has to be evaluated. We have performed a preliminary experiment which showed comparable results for signal reconstruction and fODFs.\\

\section{Conclusion}
Modeling dMRI data using NeSH produces results in downstream tasks with similar or possibly better results than established methods. It also shows promising results in the field of angular and spatial upsampling. NeSH can make use of the structural coherence in the brain, and does not rely on a prior learned on other datasets. The experiments in this paper provide an interesting avenue for modeling dMRI data, which should be further explored in future research.

\bibliography{refs}

\begin{thebibliography}{10}
\providecommand{\url}[1]{\texttt{#1}}
\providecommand{\urlprefix}{URL }
\providecommand{\doi}[1]{https://doi.org/#1}

\bibitem{ajavalidation}
Aja-Fernandez, S., Martin-Martin, C., Planchuelo-Gomez, A., Faiyaz, A., Uddin,
  N., Schifitto, G., Tiwari, A., Shigwan, S.J., Singh, R.K., Zheng, T., et~al.:
  Validation of deep learning techniques for quality augmentation in diffusion
  mri for clinical studies

\bibitem{alexander2017image}
Alexander, D.C., Zikic, D., Ghosh, A., Tanno, R., Wottschel, V., Zhang, J.,
  Kaden, E., Dyrby, T.B., Sotiropoulos, S.N., Zhang, H., et~al.: Image quality
  transfer and applications in diffusion mri. NeuroImage  \textbf{152},
  283--298 (2017)

\bibitem{basser1994estimation}
Basser, P.J., Mattiello, J., LeBihan, D.: Estimation of the effective
  self-diffusion tensor from the nmr spin echo. Journal of Magnetic Resonance,
  Series B  \textbf{103}(3),  247--254 (1994)

\bibitem{bonet2019computationalgranada}
Bonet-Carne, E., Grussu, F., Ning, L., Sepehrband, F., Tax, C.M.: Computational
  Diffusion MRI: International MICCAI Workshop, Granada, Spain, September 2018.
  Springer (2019)

\bibitem{descoteaux2007regularized}
Descoteaux, M., Angelino, E., Fitzgibbons, S., Deriche, R.: Regularized, fast,
  and robust analytical q-ball imaging. Magnetic Resonance in Medicine: An
  Official Journal of the International Society for Magnetic Resonance in
  Medicine  \textbf{58}(3),  497--510 (2007)

\bibitem{girshick2015fast}
Girshick, R.: Fast r-cnn. In: Proceedings of the IEEE international conference
  on computer vision. pp. 1440--1448 (2015)

\bibitem{mancini2022lossy}
Mancini, M., Jones, D.K., Palombo, M.: Lossy compression of multidimensional
  medical images using sinusoidal activation networks: an evaluation study
  (2022)

\bibitem{mildenhall2021nerf}
Mildenhall, B., Srinivasan, P.P., Tancik, M., Barron, J.T., Ramamoorthi, R.,
  Ng, R.: Nerf: Representing scenes as neural radiance fields for view
  synthesis. Communications of the ACM  \textbf{65}(1),  99--106 (2021)

\bibitem{ning2020cross}
Ning, L., Bonet-Carne, E., Grussu, F., Sepehrband, F., Kaden, E., Veraart, J.,
  Blumberg, S.B., Khoo, C.S., Palombo, M., Kokkinos, I., et~al.: Cross-scanner
  and cross-protocol multi-shell diffusion mri data harmonization: Algorithms
  and results. Neuroimage  \textbf{221},  117128 (2020)

\bibitem{raffelt2012apparent}
Raffelt, D., Tournier, J.D., Rose, S., Ridgway, G.R., Henderson, R., Crozier,
  S., Salvado, O., Connelly, A.: Apparent fibre density: a novel measure for
  the analysis of diffusion-weighted magnetic resonance images. Neuroimage
  \textbf{59}(4),  3976--3994 (2012)

\bibitem{sitzmann2020implicit}
Sitzmann, V., Martel, J.N.P., Bergman, A.W., Lindell, D.B., Wetzstein, G.:
  Implicit neural representations with periodic activation functions (2020)

\bibitem{tancik2020fourier}
Tancik, M., Srinivasan, P., Mildenhall, B., Fridovich-Keil, S., Raghavan, N.,
  Singhal, U., Ramamoorthi, R., Barron, J., Ng, R.: Fourier features let
  networks learn high frequency functions in low dimensional domains. Advances
  in Neural Information Processing Systems  \textbf{33},  7537--7547 (2020)

\bibitem{tax2019cross}
Tax, C.M., Grussu, F., Kaden, E., Ning, L., Rudrapatna, U., Evans, C.J.,
  St-Jean, S., Leemans, A., Koppers, S., Merhof, D., et~al.: Cross-scanner and
  cross-protocol diffusion mri data harmonisation: A benchmark database and
  evaluation of algorithms. NeuroImage  \textbf{195},  285--299 (2019)

\bibitem{tournier2007robust}
Tournier, J.D., Calamante, F., Connelly, A.: Robust determination of the fibre
  orientation distribution in diffusion mri: non-negativity constrained
  super-resolved spherical deconvolution. Neuroimage  \textbf{35}(4),
  1459--1472 (2007)

\bibitem{tournier2013determination}
Tournier, J.D., Calamante, F., Connelly, A.: Determination of the appropriate b
  value and number of gradient directions for high-angular-resolution
  diffusion-weighted imaging. NMR in Biomedicine  \textbf{26}(12),  1775--1786
  (2013)

\bibitem{tournier2010improved}
Tournier, J.D., Calamante, F., Connelly, A., et~al.: Improved probabilistic
  streamlines tractography by 2nd order integration over fibre orientation
  distributions. In: Proceedings of the international society for magnetic
  resonance in medicine. vol.~1670. Ismrm (2010)

\bibitem{tournier2019mrtrix3}
Tournier, J.D., Smith, R., Raffelt, D., Tabbara, R., Dhollander, T., Pietsch,
  M., Christiaens, D., Jeurissen, B., Yeh, C.H., Connelly, A.: Mrtrix3: A fast,
  flexible and open software framework for medical image processing and
  visualisation. Neuroimage  \textbf{202},  116137 (2019)

\bibitem{van2013wu}
Van~Essen, D.C., Smith, S.M., Barch, D.M., Behrens, T.E., Yacoub, E., Ugurbil,
  K., Consortium, W.M.H., et~al.: The wu-minn human connectome project: an
  overview. Neuroimage  \textbf{80},  62--79 (2013)

\bibitem{yu2021plenoctrees}
Yu, A., Li, R., Tancik, M., Li, H., Ng, R., Kanazawa, A.: Plenoctrees for
  real-time rendering of neural radiance fields. In: Proceedings of the
  IEEE/CVF International Conference on Computer Vision. pp. 5752--5761 (2021)

\bibitem{zhu2023deep}
Zhu, F., Guo, S., Song, L., Xu, K., Hu, J., et~al.: Deep review and analysis of
  recent nerfs. APSIPA Transactions on Signal and Information Processing
  \textbf{12}(1) (2023)

\end{thebibliography}
\bibliographystyle{splncs04}
\end{document}